\documentclass[conference]{IEEEtran}
\IEEEoverridecommandlockouts
\usepackage{cite}
\usepackage{amsmath,amssymb,amsfonts}
\usepackage{algorithmic}
\usepackage{graphicx}
\usepackage{textcomp}
\usepackage{xcolor}
\usepackage{url}
\usepackage[acronym]{glossaries}
\makeglossaries
\usepackage{soul}

\def\BibTeX{{\rm B\kern-.05em{\sc i\kern-.025em b}\kern-.08em
    T\kern-.1667em\lower.7ex\hbox{E}\kern-.125emX}}

\usepackage{array}
\usepackage{ragged2e}
\usepackage{caption}
\usepackage{booktabs}
\usepackage{lscape}
\usepackage{amsthm}
\usepackage{tikz}
\usepackage{eso-pic} 

\theoremstyle{definition}
\newtheorem{definition}{Definition}[section]
    
\begin{document}

\title{Commitment Schemes for Multi-Party Computation\\}

\author{
\IEEEauthorblockN{1\textsuperscript{st} Ioan Ionescu}
\IEEEauthorblockA{
Department of Computer Science\\
University of Bucharest\\
Bucharest, Romania\\
ioan.ionescu@unibuc.ro}
\and
\IEEEauthorblockN{2\textsuperscript{nd} Ruxandra F. Olimid}
\IEEEauthorblockA{
Department of Computer Science,\\
University of Bucharest and\\
Research Institute of the University of Bucharest (ICUB) \\
Bucharest, Romania\\
ruxandra.olimid@fmi.unibuc.ro}
}

\maketitle

\begin{abstract}

The paper presents an analysis of Commitment Schemes (CSs) used in Multi-Party Computation (MPC) protocols. While the individual properties of CSs and the guarantees offered by MPC have been widely studied in isolation, their interrelation in concrete protocols and applications remains mostly underexplored. This paper presents the relation between the two, with an emphasis on (security) properties and their impact on the upper layer MPC. In particular, we investigate how different types of CSs contribute to various MPC constructions and their relation to real-life applications of MPC. The paper can also serve as a tutorial for understanding the cryptographic interplay between CS and MPC, making it accessible to both researchers and practitioners. Our findings emphasize the importance of carefully selecting CS to meet the adversarial and functional requirements of MPC, thereby aiming for more robust and privacy-preserving cryptographic applications.

\end{abstract}

\newacronym{mpc}{MPC}{Multi-Party Computation}
\newacronym{psi}{PSI}{Private Set Intersection}
\newacronym{cs}{CS}{Commitment Scheme}
\newacronym{zk}{ZK}{Zero-Knowledge}
\newacronym{uc}{UC}{Universally Composable}
\newacronym{ppt}{PPT}{Probabilistic Polynomial-Time}
\newacronym{pvcc}{PVC}{Publicly Verifiable Covert}
\newacronym{ot}{OT}{Oblivious Transfer}

\begin{IEEEkeywords}
multi-party computation, commitment schemes, security properties, privacy
\end{IEEEkeywords}

\section{Introduction}

In recent years, \gls{mpc} protocols have become practical and are currently used in real-world applications, including distributed electronic voting, sealed-bid auctions, and privacy-preserving data analysis. A critical building block in \gls{mpc} is (many times) the cryptographic primitive called \gls{cs}, which allows one party to commit to a chosen value while keeping it hidden from other parties until its opening. \glspl{cs} are essential for fairness and security in MPC, as they bind parties to their inputs or intermediate values, allowing cheating detection (e.g., altering a secret share or bid midway) by opening the commitments~\cite{baum2016efficient, baum2024cheater}. Moreover, \glspl{cs} can also bring other features, such as guaranteeing the finalization of the protocol in case of participants' abortion~\cite{boneh2000timed,tyagi2023riggs}.
These aspects motivate the study of \glspl{cs} in \gls{mpc} protocols, with a focus on the impact the underlying (security) properties in \glspl{cs} have on the upper layer \gls{mpc}.

\subsection{Contribution}

We analyze how various types of \glspl{cs} integrate into \gls{mpc} protocols in terms of the security guarantees they enable and how these impact distinct real-world applications. We provide a relational study of representative \gls{mpc} protocols and \glspl{cs} types, highlighting their usability in relation to different functionalities and requirements. As a secondary contribution, the paper can be perceived as a tutorial on \gls{cs} and \gls{mpc}, with a focus on the fundamental properties of the two cryptographic concepts and their interrelation.

As a note, we do not aim to explore the relation between \gls{zk} and \glspl{cs} or \gls{mpc} (e.g., MPC-in-the-Head). Although \glspl{cs} are closely related to \gls{zk}-proofs and \gls{zk}-proofs have been lately in strong connection to \gls{mpc} protocols, they can be the subject of a stand-alone study and fall outside the scope of this paper.

\subsection{Outline}

The paper is structured as follows. Section \ref{sec:preliminaries} gives the preliminaries. Section~\ref{sec:cs} introduces \glspl{cs}, listing the main properties and providing a succinct classification. Section~\ref{sec:mpc} introduces \gls{mpc}, listing the main properties and exemplifying popular protocols and real-world applications, with focus on the relation with the underlying \gls{cs}. Section~\ref{sec:relation-mpc-cs} analyzes how different types of \glspl{cs} contribute to the security guarantees of \gls{mpc} protocols, defining these roles through a range of applications. Section~\ref{sec:conclusion} summarizes our findings and proposes future research directions.

\section{Preliminaries}
\label{sec:preliminaries}

\subsection{Acronyms and Notations}
\addcontentsline{toc}{section}{List of Acronyms}

\begin{center}
\begin{table}[b!]
\caption{List of acronyms}
\label{tab:acronyms}
\begin{center}
\begin{tabular}{ll}
\hline
\textbf{Acronym} & \textbf{Explanation} \\
\hline
CS & Commitment Scheme \\
MPC & Multi-Party Computation \\
OT & Oblivious Transfer \\
PPT & Probabilistic Polynomial-Time \\
PSI & Private Set Intersection \\
PVC & Publicly Verifiable Covert \\
UC & Universally Composable \\
ZK & Zero-Knowledge \\
\hline
\end{tabular}
\end{center}
\end{table}
\end{center}

Let $P=\{P_1, P_2,..., P_n\}$ denote a set of \(n\) parties participating in a protocol. We assume that party \(P_i\) inputs \(x_i\), for all $i = 1,\dots, n$, where $x_i$ is usually sensitive and thus private to $P_i$. A function that is jointly computed by the parties in $P$ is described as \(f(x_1, x_2, ..., x_n)= (y_1, y_2,...,y_n)\). We occasionally refer to a tuple using a single-letter notation, without indexes, e.g., $y = (y_1, y_2, ..., y_n)$.

Let \(k\) be the security parameter. Let $\mathcal{A}$ denote a \gls{ppt} adversary. TABLE~\ref{tab:acronyms} lists the used acronyms.

\medskip
\subsection{Adversarial Model}
\label{subsec:adv}

We consider well-established security models, both in terms of adversarial resources and behavior. We refer to the conventional \gls{ppt} adversary, whose attacks are bounded in polynomial time, in opposition to an unbounded adversary that has infinite computational power to succeed in the attacks. Note that it is common in \gls{mpc} to allow the adversary to gain control of a subset of parties participating in the protocol execution, which are referred to as \textit{corrupted}. Corruption can be established a priori of the protocol execution (\textit{static corruption}) or can adapt during the protocol execution (\textit{adaptive corruption}). In the \textit{honest-but-curious} adversarial model, all parties follow the protocol exactly as specified but may be interested in obtaining information for which they are unauthorized. In the \textit{malicious} adversarial model, malicious parties may arbitrarily deviate from the protocol execution. Reference \cite{lindell2020secure} nicely summarizes the adversarial models in \gls{mpc}.

\section{Commitment Schemes}
\label{sec:cs}

\subsection{Definition and Properties}

A \textit{\gls{cs}}~\cite{goldreich2001} is a cryptographic primitive that allows one party to commit to a value while keeping it hidden from others, with the ability to reveal it later. Informally, it can be described as a digital analogue of a sealed envelope. Once the message is placed inside, it cannot be changed (\textit{binding}), and no one can read it until the envelope is opened (\textit{hiding}).

\begin{definition}[Commitment Scheme (CS)]
A \textit{\gls{cs}} is a tuple of three algorithms $(\mathsf{Setup}, \mathsf{Commit}, \mathsf{Open})$ defined as follows: 
\begin{enumerate}
    \item $\mathsf{CK} \leftarrow \mathsf{Setup}(1^k)$: A randomized algorithm that on input the security parameter $k$ generates the public commitment key $\mathsf{CK}$.
    \item $(cs,os) \leftarrow \mathsf{Commit}(\mathsf{CK},m)$: A randomized algorithm that on input the commitment key $\mathsf{CK}$ and a message $m$ outputs a commitment string $cs$ and an opening string $os$ that the receiver will later use for opening the commitment. Once the sender commits to a certain value $m$, the value of $m$ cannot be changed (the commitment is \textit{binding}), and the receiver finds nothing about the committed value $m$ (the commitment is \textit{hiding}). 
    \item $(m,b) \leftarrow \mathsf{Open}(\mathsf{CK},com,os)$: A deterministic algorithm that on input the commitment key $\mathsf{CK}$, the commitment string $cs$ and the opening string $os$, allows the verifier to reveal the initial message $m$ and to check its validity ($b = 1$ if the commitment verifies correctly and $b=0$ otherwise).
    
\end{enumerate}
\end{definition}

The \textit{hiding} and \textit{binding} properties are essential to \glspl{cs}, but \glspl{cs} might possess other properties too. TABLE~\ref{tab:cs-properties} summarizes the most relevant properties a \gls{cs} can present. We discuss these properties in relation to their impact on the \gls{mpc} that employs a \gls{cs} in Section \ref{sec:relation-mpc-cs}.

\begin{table}[t!]
\centering
\caption{Properties of Commitment Schemes}
\label{tab:cs-properties}
\renewcommand{\arraystretch}{1.3}
\setlength{\tabcolsep}{6pt} 
\begin{tabular}{p{2cm}p{6cm}}
\hline
\textbf{Property} & \textbf{Informal Definition} \\
\hline
\textit{Hiding} & The committed value remains hiden until the opening phase~\cite{goldreich2001} 
\\
\textit{Binding} & The committer cannot change the committed value or open it in two different ways~\cite{goldreich2001} \\

\textit{Homomorphism} & The commitment supports algebraic operations (e.g., additive or multiplicative) on committed values, without opening them~\cite{pedersen_commitment, committed_mpc,catalano2005multiparty,damgaard2002perfect} \\
\textit{Non-malleability} & Prevents generating valid commitments for related messages (e.g., predictable transformation of the original message)~\cite{goldreich2001,damgaard2002perfect} \\
\textit{Interactivity} & Indicates whether the scheme is interactive (multi-round) or non-interactive (one-round)~\cite{goldreich2001} \\
\textit{Succinctness} & Commitments and openings are small, aiding efficiency and scalability (ideally logarithmic in the size of the message)~\cite{mpc_friendly_commitments} \\
\textit{Equivocability} & In trapdoor setups, a committer with special knowledge can open the commitment to any value (i.e., the committer can change his/her mind)~\cite{goldreich2001} \\
\textit{Extractability} & Using a trapdoor, a trusted party can extract the message from the commitment (i.e., one can efficiently compute the message from the commitment)~\cite{goldreich2001} \\
\textit{Timed Commitment} & Commitment can be forcibly opened after a delay (using time-lock puzzles or delay functions)~\cite{boneh2000timed}\\
\textit{Public \mbox{Verifiability}} & Anyone can verify the validity of a commitment, without interaction~\cite{baum2014publicly, agrawal2021mpc}. In the settings of \gls{pvcc} security model, cheaters aim to prevent being discovered, as they face a reputational harm \cite{mpc_friendly_commitments}. \\
\hline
\textit{Perfect} & An unbounded party breaks one property with probability zero (e.g., \textit{perfect hiding} means that an unbounded receiver gets zero information about the message, and \textit{perfect binding} means that an unbounded committer can change the message with probability zero\cite{damgaard2002perfect})\\
\textit{Adaptability} & The scheme remains secure even if the adversary chooses inputs adaptively, based on previously seen commitments~\cite{byali2017fast,damgaard2002perfect} \\
\textit{\gls{uc}} & Security is preserved under concurrent executions of an unbounded number of scheme copies, adaptive corruption, and asynchronous settings~\cite{canetti2001uc,damgaard2002perfect} \\
\textit{Post-quantum} \textit{\mbox{resistance}} & The ability of a \gls{cs} to stand against quantum attacks~\cite{ajtai1996generating}. \\
\hline
\end{tabular}
\end{table}

\subsection{Classification}
\label{subsec:cs_classif}

\glspl{cs} can be classified according to several criteria. 
In particular, \glspl{cs} can be classified according to the key properties that determine their functionality, efficiency, and suitability in cryptographic protocols, particularly in secure \glspl{mpc}. These properties might include interactivity, support for algebraic operations (e.g., homomorphism), time-based behavior, verification model, extractability, and others, as listed in TABLE~\ref{tab:cs-properties}.

Concerning the type of the committed value, \textit{bit} commitments (i.e., commit to a single bit) \cite{gmw,chaum1988multiparty}, \textit{vector} commitments (i.e., commit to an ordered set of values and open it at specific positions) \cite{catalano2013vector,pathak2025vector}, \textit{polynomial} commitments (i.e., commit to a polynomial and open it at evaluations of the polynomial) \cite{pathak2025vector,private_polynomial_commitments} have been successfully used in \gls{mpc}.

Concerning the adversarial model, there is a clear distinction between the \textit{computational} fulfillment of a given property and its \textit{perfect} satisfiability. Lately, there has been an increase in interest in post-quantum candidate primitives, including post-quantum candidates \gls{cs}.
We refer to some of the security-modeling-related notions in the second part of TABLE~\ref{tab:cs-properties}.

\section{Multi-Party Computation Protocols}
\label{sec:mpc}

\subsection{Definition and Properties}

\textit{MPC} is a cryptographic protocol that allows multiple parties to jointly compute a function on their private inputs. The parties learn the result of the function and (ideally) nothing else. In particular, parties do not find other parties' inputs. 
Although there is sometimes a distinction in terminology between secure MPC and \gls{mpc}~\cite{evans2018pragmaticmpc}, for simplicity, we refer to secure \gls{mpc} as \gls{mpc}~\cite{evans2018pragmaticmpc,lindell2020secure}.

\begin{definition}[Secure Multi-Party Computation (MPC)]
A \textit{\gls{mpc} Protocol} for a \(n\)-ary function $f$ is a cryptographic protocol run between \(n\) parties \(P_1, P_2,..., P_n\), each holding a corresponding secret value \(x_1, x_2,..., x_n\) so that the parties jointly compute the value $f(x_1, x_2,..., x_n)$, in such a way that no information about the input values can be inferred from the result (\textit{privacy}), and that the result is \textit{correct}, even when a certain fraction of the parties are corrupted. 
\end{definition}

While the main objectives of an \gls{mpc} protocol are to protect the \textit{privacy} of inputs and to ensure the \textit{correctness} of the output, even in the presence of corrupted parties, similar to \glspl{cs}, \gls{mpc} might present other properties too. TABLE~\ref{tab:mpc-properties} provides a summary of the key properties for \gls{mpc}. As a novelty, we present the \gls{mpc} properties in relation to the underlying \gls{cs} and the benefits it brings. Note that some relations are immediate (e.g., \textit{succintness}), while others are not immediate (e.g., \textit{fairness} of \gls{mpc} from \textit{timed commitments}).

\begin{table}[t!]
\centering
\caption{Properties of MPC protocols}
\label{tab:mpc-properties} 
\renewcommand{\arraystretch}{1.3}
\setlength{\tabcolsep}{6pt} 
\begin{tabular}{p{2cm}p{6cm}}
\hline
\textbf{Property} & \textbf{Informal Definition} \\
\hline
\textit{Correctness} (\textit{Robustness}) & The protocol computes the correct output as if executed honestly. In malicious settings, \gls{cs}s enforce input consistency and allow later verification~\cite{committed_mpc} \\
\textit{Privacy} & Ensures no party learns more than their input and the output. \gls{cs}s help hide intermediate values and prevent input leakage during setup~\cite{private_polynomial_commitments} \\
\textit{Fairness} & All parties learn the output or none does. Timed or verifiable \gls{cs}s enforce fairness by enabling time-locked openings or detecting misbehavior~\cite{naor1999privacy} \\
\textit{(Public) Verifiability} & Allows parties (or external auditors) to check output correctness based on committed inputs. Publicly verifiable \gls{cs}s support third-party verification~\cite{baum2024cheater} \\
\textit{(Public) Auditability} & Extends verifiability with traceability over protocol steps, allowing one party that only accesses the public transcript of the computation to check the output correctness \cite{baum2014publicly}. Extractable or \gls{uc}-secure \gls{cs}s enable retrospective audits or simulations~\cite{canetti2001uc} \\
\textit{Accountability} & An honest party can convince a third party that another party cheated in the computation \cite{baum2016efficient} \\
\textit{Dynamicity} & The protocol supports dynamic changes of the group of parties taking part in the protocol execution (join/leave the group) without compromising security or correctness~\cite{wenxuan2022admpc} \\
\textit{Asynchronism} & Correctness and security are preserved even with delayed, unordered, or uneven message delivery. No timing assumptions are required, making it suitable for unpredictable networks~\cite{wenxuan2022admpc} \\
\textit{Succinctness} & Refers to the compactness of exchanged messages and commitments. Improves scalability in bandwidth-constrained settings~\cite{private_polynomial_commitments} \\
\textit{Computational Efficiency} & Captures the per-party cost (e.g., encryption, exponentiation). Efficient protocols minimize expensive cryptographic operations~\cite{committed_mpc} \\
\textit{Interactivity} & Describes the number of interaction rounds. Non-interaction or low-interaction protocols (e.g., constant-round) offer better performance and fault tolerance~\cite{canetti2001uc, damgaard2012spdz} \\
\hline
\end{tabular}
\end{table}

The most well-known example of MPC is Yao's millionaire problem \cite{yao_secure_computation}, an instance of \gls{mpc} with two parties (\(n=2\)) representing two millionaires who want to decide who is richer without revealing their wealth to each other. Many other \gls{mpc} protocols have been proposed since. We exemplify with the GMW~\cite{gmw} and SPDZ~\cite{damgaard2012spdz} protocols, as representatives for shaping the evolution of the field.

\subsection{Classification and Applications of MPC}

Similar to \gls{cs}s, \gls{mpc} protocols can be classified according to their key properties, listed in TABLE~\ref{tab:mpc-properties}. It is essential to mention that efficiency in \gls{mpc} is often analyzed through several dimensions, including communication cost, round complexity, and succinctness. While these aspects are interrelated, for clarity, we list them separately.
In terms of security modeling, the traditional distinction between bounded and unbounded adversaries remains in place, as already discussed in Sections \ref{subsec:adv} and \ref{subsec:cs_classif}. We again highlight the adversarial model, with strong capabilities in the adaptable malicious model (see Section \ref{subsec:adv})

We briefly exemplify next some areas of applicability for \gls{mpc}, highlighting the role of underlying \gls{cs}.

\paragraph{Electronic Voting}
\gls{mpc} enables verifiable and private voting; tallies are computed without revealing individual votes. \gls{cs} preserve vote secrecy until opening~\cite{kiayias2015e2e}.

\paragraph{(Sealed-bid) Auctions}
\gls{mpc} computes winners without revealing losing bids. \gls{cs} prevent bid alteration. Timed commitments ensure fairness and enforce deadlines~\cite{naor1999privacy}.

\paragraph{\gls{psi}}
Parties compute the set intersection of their inputs privately, without revealing any information beyond the result. In \gls{mpc}-based \gls{psi} protocols, commitments can prevent input substitution and improve integrity~\cite{freedman2004efficient,morales2023private}.

\paragraph{Threshold Cryptography}
\gls{mpc} supports distributed, threshold encryption and signature schemes. Commitments ensure input consistency during distributed key generation and partial decryption~\cite{shamir1979share}.

\paragraph{Federated Machine Learning}
\gls{mpc} enables secure inference or training on encrypted and distributed data, without revealing either the data or the model. \glspl{cs} enforce correct model use (e.g., structure, weights, preprocessing) and enable auditing~\cite{mohassel2017secureml}.

\paragraph{Federated Analysis}
Authorized entities can leverage \gls{mpc} to perform aggregate computations over sensitive datasets (e.g., tax records, census data, health information) held by different entities. This enables privacy-preserving statistics while preserving data sovereignty. Such applications typically require strong input validity and consistency guarantees, often enforced through \glspl{cs} and \gls{zk} proofs~\cite{bogdanov2014privacy}.

\paragraph{Others}
\gls{mpc} is also used in secure financial benchmarking, private genomic computation (e.g., computing disease risks based on sensitive genetic data), database applications, and key management applications, among others, with direct usability in real-life applications \cite{archer2018keys}. These examples illustrate the wide applicability of \gls{mpc} beyond theoretical cryptographic use cases.

\gls{mpc} protocols are utilized across various areas, each with its own functional and security requirements. To illustrate, voting protocols prioritize public verifiability and input privacy, whereas auction systems focus on fairness and bid consistency. These applications often require cryptographic tools that enforce input validity, enable public audits, or allow verifiable execution without revealing sensitive data. In this respect, \glspl{cs} play a critical role by ensuring that parties cannot alter their inputs once they are committed to, while preserving secrecy until the opening phase. The next section provides a detailed discussion of the security requirements that \gls{mpc} protocols must fulfill and how \glspl{cs} contribute to achieving these properties in various scenarios.

\section{Discussion}
\label{sec:relation-mpc-cs}

We delve further into a relational study of representative \gls{mpc} protocols and their underlying \glspl{cs}, keeping in mind that using \glspl{cs} in \gls{mpc} protocols is crucial for enforcing various security properties across different contexts.

In SPDZ~\cite{damgaard2012spdz}, an actively secure \gls{mpc} with dishonest majority, commitment techniques help to grant integrity and privacy guarantees. Each party effectively commits to its secret-shared inputs and intermediate values (e.g., via explicit cryptographic commitments) so that any attempt to cheat by altering a value can be detected in a final consistency check~\cite{baum2014publicly}. The binding property of the underlying \gls{cs} ensures that parties cannot change their values once committed to, while the hiding property preserves the privacy of these values throughout the computation. Extensions of SPDZ employ Pedersen Commitments~\cite{pedersen_commitment} (a computationally binding and information-theoretically hiding scheme) to achieve public verifiability: every input, intermediate share, and output is posted as a commitment, allowing an external auditor to verify correctness after execution~\cite{baum2014publicly}. In summary, commitments in SPDZ provide active security with abort – no adversary can cause an incorrect result to be accepted – and even enable fairness and auditability in variants where misbehavior is detected or prevented. 

In the classic GMW~\cite{gmw} protocol for circuit-based \gls{mpc}, commitments can provide security and fairness guarantees against malicious attacks. Originally, GMW assumes honest-but-curious adversaries. To harshen it against adversaries that might deviate from the protocol execution, a common technique is to have the parties commit to all their random coin tosses or secret shares before executing the circuit computation~\cite{scholl_et_al2022}. Thus, the binding property prevents cheating, as any deviation would contradict a prior commitment. A detected inconsistency leads to abortion, and the given party is identified as malicious. Hence, the use of commitments in GMW facilitates correctness (i.e., the output reflects honest inputs) and partial fairness.

Many efficient \gls{mpc} applications rely on simple hash-based \gls{cs}s to enforce input or randomness commitments. A hash function \(H\) can serve as a commitment by computing, for instance, \(Com(x, r) = H(x || r)\), which is binding and hiding under cryptographic assumptions~\cite{cryptoeprint_2023/030}. Such hash-based commitments are lightweight and common in protocols such as coin-tossing and auctions. For example, in a two-party coin flip, one party commits to a random bit by sending its hash. The other party then chooses a bit, after which the first party opens the hash commitment. This ensures that the first party cannot change its chosen bit after seeing the other’s choice, achieving bias resistance (the outcome is cryptographically bound to the committed value) and hiding the bit until its opening. Similarly, in sealed-bid auctions, all bidders submit commitments (often hash-based) to their bids before any bid is revealed, guaranteeing that no bidder can alter or retract them later. The hiding property keeps bids secret, while binding ensures integrity, as the auction outcome is computed from the originally committed bids.

Timed commitments appear in applications such as auctions and coin-flipping to achieve stronger notions of fairness. In an auction, if a bidder fails to reveal their committed bid by the deadline, the protocol (or the other parties) can still recover the hidden bid after a specific period, preventing a dishonest party from influencing the auction outcome \cite{boneh2000timed,tyagi2023riggs}. This achieves fairness: no one can gain an advantage by refusing to open a commitment. The two-party coin flipping is similar. Timed commitments ensure that even if one party attempts to abort upon seeing an unfavorable partial outcome, the other party can eventually force the opening and obtain the result. The security guarantee here is fairness (or termination) in the sense that no party can indefinitely block the protocol execution \cite{boneh2000timed}. Consequently, timed commitments extend the fundamental guarantees with a temporal dimension, which is crucial for fairness in \gls{mpc} competitive scenarios.

\begin{table*}[t!]
\centering
\caption{Comparison of Commitment Schemes in MPC Protocols}
\label{tab:commitment-comparison}
\renewcommand{\arraystretch}{1.3}
\begin{tabular}{p{3cm}p{2.5cm}p{3.5cm}p{3.5cm}p{3.5cm}}
\hline
\textbf{Property of the \gls{cs}} & \textbf{MPC Protocol} & \textbf{Role of CS in MPC} & \textbf{Key Properties} & \textbf{Application} \\
\hline

Homomorphic & SPDZ \cite{damgaard2012spdz} & Input masking & Perfect hiding, Computational binding, Homomorphic & Secure computation, Voting \\
Homomorphic, Publicly Verifiable & Committed MPC \cite{committed_mpc} & Define a secret-sharing based MPC &  Secure MPC against Malicious dishonest majority & Committed MPC, Committed OT\\

Polynomial & Custom MPSI \cite{private_polynomial_commitments} & Enforce correctness of polynomial evaluations & Binding, Completeness & Verifiable secret sharing, MPC over polynomials \\

Publicly Verifiable & Cheater identification \cite{baum2024cheater, cunningham2017catching, agrawal2021mpc,spini2016cheater} & Detect and identify misbehavior in output phase & Public verifiability, Binding, Auditability & Malicious-secure MPC, Voting \\

\gls{pvcc} & \gls{pvcc} -secure MPC \cite{mpc_friendly_commitments} & Detect and identify cvert misbehavior & Public verifiability & Reputation-preserving MPC \\

Non-interactive (Hash-based) & GMW variant \cite{brakerski2017four} & Prevention of input manipulation & Computational hiding, Binding, Non-malleability, Adaptive security & General \gls{mpc} without setup \\

Timed Commitments & Privacy Preserving Auctions~\cite{naor1999privacy,boneh2000timed} & Time-locked reveal phase & 
Fairness & Auctions, Coin flipping \\

Non-malleable & General \gls{mpc} \cite{byali2017fast} & Prevent adversary from committing to related value & Non-malleability, Binding, \gls{uc}-security & \gls{uc}-secure \gls{mpc}, \gls{ot}, Composable protocols \\

Extractable & \gls{uc}-secure \gls{mpc}~\cite{canetti2001uc} & Supports simulation by enabling extraction of committed inputs in the security proof & Extractability, Simulation-friendliness & \gls{uc}-security protocols \\

\hline
\end{tabular}
\end{table*}

Secure \gls{psi} protocols also use \gls{cs}s to enforce input and output security properties. In \gls{psi}, two or more parties learn the intersection of their private sets and nothing else. A malicious party might attempt to cheat by adaptively inserting extra elements or by providing false input. To prevent this, some \gls{psi} constructions have each participant commit to their input set elements (or to a Merkle tree root of all elements in the set) at the start of the protocol. Because of the binding property, once a party commits to the inputs, he/she cannot change them. This guarantees input consistency throughout the protocol \cite{cryptoeprint:2024/789}. For example, a party commits to each item in a set (e.g., using a vector \gls{cs}) and only opens the commitment corresponding to elements in the intersection to prove their validity. As such, the commitments remain hidden until the appropriate opening phase, and the elements outside of the revealed intersection remain hidden. Thus, using commitments in \gls{psi} ensures correctness (the computed intersection is sound, derived from genuine inputs) and privacy (no additional information about non-intersecting items leaks). \gls{cs} can also contribute to the \gls{psi} fairness if the parties exchange commitments but only open them at the end after fulfilling the required steps. This ensures that one party cannot learn the intersection and then abort without revealing its contributions \cite{cryptoeprint:2024/789}.

Finally, commitments enable auditable \gls{mpc}, where an external entity can verify the correctness of a computation after it has been completed. In an auditable \gls{mpc} scenario (e.g., in electronic voting or verifiable auctions), the participants commit to all their inputs and every intermediate computation value during the protocol, producing a public transcript of commitments \cite{baum2024cheater}. Along with these, they provide proofs or later openings for each step. Because each committed value is bound to the actual secret used in the computation, any deviation from the protocol would result in an inconsistent set of opened commitments, which an auditor would notice. Moreover, suppose commitments are homomorphic (e.g., Pedersen commitments \cite{pedersen_commitment}). In that case, an auditor can verify aggregated relations (with respect to addition or multiplication on secret-shared values) without needing to examine the private inputs. The security guarantee afforded by this approach is public verifiability or accountability: even if all the computing parties were corrupt, anyone reviewing the commitment transcript could be confident that either the computation was correct or a detectable inconsistency would be present. Notably, this does not compromise privacy: the commitments are hiding, so the auditor learns nothing about individual inputs beyond what the output reveals. To illustrate, Baum et al. \cite{baum2014publicly} apply this technique to create auditable versions of the SPDZ protocol, where the online phase utilizes a public commitment bulletin board.

The mentioned scenarios illustrate how specific properties of \gls{cs}s map to the needs of different \gls{mpc} applications. From facilitating correctness and privacy in general \gls{mpc} to achieving fairness in coin flips and auctions, enforcing input honesty in \gls{psi}, and providing public accountability in auditable \gls{mpc}, commitments serve as a tool to bridge cryptographic guarantees with the desired security needs in \gls{mpc}. In this respect, TABLE~\ref{tab:commitment-comparison} provides a comparative overview of representative \gls{cs} types, highlighting their main properties, usage in \gls{mpc} protocols, and application domains. This classification supports the discussion presented in the current section. Several of the \gls{cs} presented in TABLE~\ref{tab:commitment-comparison} have practical implementations. For instance, homomorphic commitments in the SPDZ protocol are available in the MP-SPDZ library~\cite{mp_spdz}, and polynomial commitments in open-source tools such as \cite{arkworks_polycommit}. Other schemes, such as the timed or extractable commitments in \cite{naor1999privacy,boneh2000timed}, have been implemented only as proof-of-concept or remain largely theoretical.

In addition to established protocols such as SPDZ~\cite{damgaard2012spdz} and GMW~\cite{gmw}, several recent constructions further illustrate the flexibility of \glspl{cs} in \gls{mpc}. The Committed MPC protocol~\cite{committed_mpc} employs a two-party homomorphic \gls{cs} to ensure secure input consistency and output correctness. Private polynomial commitments~\cite{private_polynomial_commitments} extend these guarantees by preserving the privacy of evaluation points in polynomial-based \gls{mpc}. Meanwhile, the scheme introduced in~\cite{mpc_friendly_commitments} enables \gls{pvcc} security, offering a trade-off between efficiency and auditability. These approaches reinforce the idea that \glspl{cs} are central to ongoing advances in \gls{mpc}.

In summary, \glspl{cs} are a foundational cryptographic tool in \gls{mpc}, and their role becomes even more critical in malicious adversarial settings.

\section{Current limitations and open problems}

While efficiency is listed as a key property of \glspl{cs} in TABLE~\ref{tab:cs-properties}, it is worth investigating how different \gls{cs} constructions affect real-world \gls{mpc} performance. For example, homomorphic commitments, such as Pedersen~\cite{pedersen_commitment}, are widely considered to be efficient in arithmetic circuits due to their reliance on basic group operations. However, in \gls{mpc} protocols, they may incur additional communication costs. In particular, if many commitments have to be computed, the large dimension of the underlying field can become a bottleneck. Inefficient handling of commitments and delays in synchronization impact the throughput and latency of maliciously secure \gls{mpc}~\cite{damgaard2012spdz,rushing_spdz}. Optimizing \glspl{cs} is significant for the scalability of upper-layer systems, particularly for large networks or resource-constrained environments. A systematic study in this respect remains a direction for future work.

Moreover, integrating properties such as post-quantum resistance and \gls{uc} security into large-scale \gls{mpc} frameworks presents non-trivial challenges. Post-quantum \glspl{cs}, such as those based on the Learning with Errors (LWE)~\cite{brakerski2017four,lyubashevsky2021lattices,badrinarayanan2021concurrent} or Short Integer Solutions (SIS)~\cite{ajtai1996lattices} assumptions, typically incur higher costs compared to classical constructions. Similarly, achieving \gls{uc} security often increases protocol complexity and execution time. \gls{uc}-secure \glspl{cs} by themselves proved to introduce enough complexity to introduce bugs in the security analysis \cite{blazy2013uccommitments}. Using them as building blocks naturally makes \gls{mpc} protocols inherit existing vulnerabilities, too. Similarly, the complexity of the underlying \gls{cs}s usually introduces some lower boundaries for the corresponding \gls{mpc} protocol. Considerable work, including \cite{garay2014uccommitments,frederiksen2015homomorphicuc}, has been performed during the years on the complexity of \gls{uc} commitments with specific properties valuable in \gls{mpc}, such as homomorphism. Nonetheless, more efficient designs and implementations of such \gls{cs}s remain a key open problem for scalable \gls{mpc} systems that aim to offer long-term security and adaptability to unexplored challenges.

Finally, an open challenge remains the design of dynamic and efficient \gls{mpc} that are suitable for collaborative scenarios such as federated analytics. Despite the development of \gls{mpc} in recent years, the dynamic case in which parties can join or leave at any point remains partially unsolved, particularly in terms of real-world usability.

\section{Conclusion}
\label{sec:conclusion}

The paper presented an analysis of how different types of \glspl{cs} contribute to the various requirements of \gls{mpc}. In particular, we highlighted the main properties of \glspl{cs} (e.g., binding, hiding, homomorphic, non-malleability, extractability) and discussed their relation in enforcing similar properties in \gls{mpc} (e.g., correctness, privacy, fairness,  dynamicity) across various real-world applications.
By simply selecting \glspl{cs} with specific properties, one can facilitate \gls{mpc} constructions to meet both functional and adversarial requirements.

Several research directions remain open for exploration. An example includes the study of post-quantum \glspl{cs} and their integration in \gls{mpc} protocols, especially in settings that require long-term confidentiality preservation. A second direction is the experimental evaluation of commitment-based \gls{mpc} implementations, particularly in terms of scalability and performance in large networks or resource-constrained environments. Finally, an open challenge remains the design of dynamic and efficient \gls{mpc} that are suitable for collaborative scenarios such as federated analytics.

\section*{Acknowledgment}
This work was supported by a grant of the Ministry of Research, Innovation and Digitalization, CNCS/CCCDI-UEFISCDI, project number ERANET-CHISTERA-IV-PATTERN, within PNCDI IV.

\bibliographystyle{IEEEtran}
\bibliography{bibliography}

\end{document}